\documentclass[prl,aps,twocolumn]{revtex4}

\def\<{\langle} \def\>{\rangle}

\newcommand{\be}{\begin{equation}}

\newcommand{\bea}{\begin{eqnarray}}
\newcommand{\eea}{\end{eqnarray}}

\usepackage{graphics}
\usepackage[]{color}

\begin{document}

\definecolor{magenta}{rgb}{1,0,1}
\definecolor{555}{rgb}{0.5,0.5,0.5}
\definecolor{999}{rgb}{0.9,0.9,0.9}
\definecolor{0095}{rgb}{0.,0.,0.95}

\title{Fast and secure key distribution using mesoscopic coherent states of light.}
\author{Geraldo A. Barbosa\footnote{geraldoabarbosa@hotmail.com}}
\affiliation{ Northwestern University, Department of Electrical and Computer Engineering, \\ 2145 N. Sheridan Road,
Evanston, IL 60208-3118, US}

\date{This version: April 27, 2004.}


\begin{abstract}

This work shows how two parties A and B can securely share sequences of random bits at optical speeds. A and B possess
true-random physical sources and exchange random bits
 by using a random sequence received to cipher the following one to be sent.
 A starting shared secret key is used and the method can be described as an unlimited one-time-pad extender.
 It is demonstrated that the minimum probability of
error in signal determination  by the eavesdropper can be set arbitrarily
close to the pure guessing level. Being based on the $M$-ry encryption protocol this method also allows
for optical amplification without security degradation, offering practical advantages over
the BB84 protocol for key distribution.

\end{abstract}

\maketitle

\section{Introduction}
\label{introduction}

Physical cryptography can create schemes providing
 two users, at distinct locations, with on-demand copies of a secure sequence of random bits
of arbitrary length and at fast rates. These schemes could be of
high value for commercial systems operating over long distances.
Based on physical laws instead of mathematical complexities,
communication with perfect secrecy could be guaranteed over an
insecure channel in Vernam's sense of a one-time-pad. Technology
advances, therefore, such as enhanced computational power, should
not affect the security of these schemes. The BB84 quantum
protocol for key distribution~\cite{bennett-brassard}, the
paradigm among protocols of this type, has been used in short
distance applications \cite{alamos} but not in long distance
networks. One fundamental reason is that the same no-cloning
theorem that guarantees its security level prohibits the signal
amplification necessary in long-haul communication links. No
practical alternative quantum scheme using quantum repeaters or
entangled states has yet been proposed although theoretical
studies exist \cite{qrepeater}. Other practical impediments are
the slow speed of the photon sources and the large recovery time
of single photon detectors.

Recently,  Yuen proposed \cite{private}  a ciphering scheme
utilizing an $M$-ry bases system that was implemented for data
encryption \cite{ykb,qcmc02}. Ref. \cite{ykb} introduces the
$M$-ry scheme and presents its first prototype-level
implementation. Ref. \cite{qcmc02} gives a more complete
description of these systems.
 Basically, in
these cryptographic prototypes, known as $\alpha \eta$ ($\alpha$
standing for coherence and $\eta$ for efficiency) systems, the
quantum noise inherent to coherent states forces different
measurement results for the eavesdropper and the legitimate users
that use a shared key in their measurements.
 This
noise will increase the observational uncertainty preponderantly
for the eavesdropper, Eve (E), rather than Alice (A) and Bob (B),
the legitimate users.
 Although this noise is irreducible by nature to all observers, the knowledge of the key allows A and B to
 discard this noise while it points to the correct information.
  The very simple idea behind this is that, for each bit, the noise inherent to the stated and generated at the emitter
  is distributed without control among the output ports in
Eve's measurement apparatus while A and B use the key to select a single output port where the noise does not practically
affect bit readings.


In this work a {\em key distribution} method is presented that
also utilizes an $M$-ry bases ciphering scheme similar to the one
described in \cite{ykb} for the purpose of {\em data encryption}.
Each basis in the $M$-ry set of bases defines {\em two} orthogonal
states whereas these bases are {\em non-orthogonal} among
themselves \cite{bases}. In these schemes a starting shared secret
key is assumed between A and B.
 The phrase ``key distribution'' is being used here to denote that one party sends to the other   random bits
created by a truly random physical process. The exchange of random bits between A and B is done in such a way that
the quantum noise of the light does not  allow E to obtain the final random sequence shared by A and B.
In contrast, a classical key expansion method could mean a process to generate mathematically
--e.g., by one-way functions-- two identical sets of random bits, one for each user,  from a set of shared starting bits.
 Stream-ciphers, for example,  generate a stream of pseudo-random bits from a starting key. However, this deterministic process produces correlations
that can be detected by the eavesdropper. Known-plaintext attacks
are particularly useful to exploit these correlations in classical
cryptography. In the $M$-ry data encryption scheme, a stream
cipher is used to generate the running key and the quantum noise
of light protects against the correlations.

The key distribution method presented in this work uses
 physical sources to guarantee the true randomness of signals.
 As in the data encryption scheme, the quantum noise of
light provides the ultimate basic protection against signal
identification. After presenting a set of basic conditions to be
obeyed by the system and the physical resources needed for A and
B, the key distribution protocol will be described step-by-step.
Each step will be followed by a brief description of its possible
implementation using the described physical resources. Very
briefly, these protocol steps describe how A and B succeed in
sending new random sequences of bits from one to the other
securely through judicious use of the quantum noise of light. This
security is achieved by using a correct combination of average
number of photons per bit and number of ciphering bases $M$, as
will be shown. The bit encoding mechanism and the associated
physical protection will then be discussed and a measure of the
minimum probability of error forced by the system on the
eavesdropper will be achieved. After showing that the system obeys
the established conditions, conclusions will be presented.
\section{Basic conditions}
First, a set of conditions will be defined to specify the
boundaries within which the problem has to be solved:\\
I) The eavesdropper is allowed to have full access to the random
signal sequence being generated.
 Granting full access to the signal should be understood as similar to an opaque attack
or,  giving Eve a perfect quantum copy of the signal sequence.
Anyway, Eve does not need to subtly tap the channel to obtain the
signals. Eve could perform arbitrary measurements on this sequence
or she could generate as many realistic (imperfect) copies as she
wants. The unrestricted access to the signal sequence is the best
(idealized) possible condition given to the eavesdropper.
\\
II) Eve samples all signals near the source, such that energy loss
does not affect her data.\\

It will be initially assumed that all parties have similar
detectors; the simplest possible assumption would be of noiseless
detectors with efficiency 1.  However, it will be shown
 that although the eavesdropper needs high signal resolution to
 distinguish between two closest bases in the $M$-ry
 system and precision to identify a sent basis,  the legitimate users do not
need such strict conditions. Therefore, the detectors used by  A
and B can be less efficient. It will be  demonstrated how one can
implement a secure key distribution system where the minimum
bit-by-bit eavesdropping probability of error can be arbitrarily
set at the pure guessing value of 1/2.

As will be shown, the protection of this scheme does not rely on an  intrusion detection mechanism, but instead on the
measurement advantage enjoyed by A and B over the eavesdropper, thanks to the knowledge of the key.

\section{The key distribution protocol}

\subsection{Basic physical resources}

\begin{figure}
\centerline{\scalebox{0.4}{\includegraphics{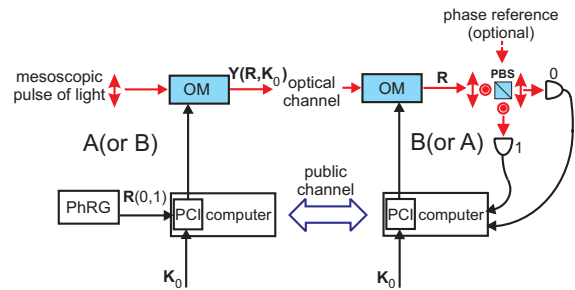}}}
\caption{\label{fig1} Basic scheme for key distribution. An
emitter-to-receiver part is shown. OM is an optical modulator.
PhRG is a {\em physical} random generator, PCI is a PC controlled
interface card. ${\bf K}_0$ is the starting running key. ${\bf
Y}({\bf R},{\bf K}_0)$ is the ciphered ${\bf R}$. A polarization
beam splitter PBS followed by two detectors constitute the
detection system in the case of polarized signals. In the case of
phase modulation,  a phase sensitive detection system should be
used.}
\end{figure}

The basic resources necessary for implementation of this
 key distribution protocol are sketched  in Fig. \ref{fig1}.
Two stations, A and B, are represented where the optical channel can be either  free space or a fiber channel.
Both sides have identical resources to operate as emitter or receiver \cite{one way}.
The OM's  are optical modulator systems performing polarization or phase modulation on {\em mesoscopic} coherent pulses of light.
Each party also possesses a fast speed {\em physical} random
generator (PhRG) that produces binary outputs ${\bf R}$. PCI is a
PC controlled interface card that can generate $M$ voltage levels.
PBS is a polarizing beam splitter that is followed by two
detectors designated by 0 and 1. This detection system can be used
for polarized light signals. In the case of phase modulation the
detection system should be modified accordingly.

\subsection{The protocol}

Each of the seven protocol steps will be stated briefly  (in
italics) and for a more complete illustration of the scheme, a
short description of one possible way to implement each step will
follow:

{{\bf 1.} \em Parties A and B share an initial secret random sequence
(of length $L_0$) of bits ${\bf K}_0$.}\\
How A and B will share this initial sequence is, of course, an
important matter. Although current cryptography can provide enough
security for sharing the short sequence ${\bf K}_0$ at this
moment, it may be vulnerable to the evolution of computational
power. Just as an example, ${\bf K}_0$ could be obtained in a
secure way within a few years through the {\em slow} but proved
secure BB84 key distribution system. The use of satellites to
distribute quantum keys have been under study (See Ref.
\cite{quantph03051} and references therein) and developments in
this direction can be expected to produce fruits in a near future.
Unless proved otherwise, the expected rates of this quantum key
distribution will be low. The scheme studied in this paper aims to
create a {\em fast} distribution rate once a short sequence ${\bf
K}_0$ have been obtained - even through a slow process.

{{\bf 2.} \em Party A generates a sequence (of length $L_0$) of
true random bits {\bf  R}.} \\
This sequence of  bits ${\bf R}$ can be obtained from the binary
output of the {\em physical} random generator (PhRG) as voltages
$V_R=V_+$ or $V_ {-}$ that are going to be
  associated with bits 0 and 1.
A possible visualization of such a process could be the voltage outputs
$V_i\:(i=1,2,\cdots)$ within a short time window $\Delta t$,
around $t_i$, produced by a fast light detector,
shot-noise limited, illuminated by a coherent light beam.
The {\em sign} of these pulses,
$sign_i=(V_i-\overline{V_i})/(|V_i-\overline{V_i}|)$,
where $\overline{V_i}$ is the average pulse voltage,
will feed a binary voltage source to provide the random
bit sequence ${\bf R}$ \cite{zbinden}.

{{\bf 3.} \em A sends to B the random sequence ${\bf R}\:(\equiv
{\bf R}_1)$ of length $L_0$ in blocks of size $K_M$. Ciphering
each of these blocks uses $K_M$ bits from ${\bf K}_0$. The number
of blocks to be ciphered in $L_0$ is $L_0/K_M$.
A coherent state carrier is used with intensity $\langle n \rangle/$bit}.\\
In order to generate each cipher basis $k \:(=0,1,\cdots M-1)$,
$K_M \:(=\log_2 M)$ bits are used from the random sequence of bits
${\bf K}_0$ (e.g, $k=b(K_M) 2^{M-1}+b(K_{M-1})2^{M-2}+\cdots
b(K_1) 2^0$). In other words, each $k$ basis of the $M=2^{K_M}$
set will be randomly defined by $K_M$ bits taken from ${\bf K}_0$.
Each $k$ will be used to cipher a block sequence of size $K_M$
from ${\bf R}_1$. Ciphering
 ${\bf R}_1$ in blocks of size $K_M$
keeps the length of the transmitted bits constant and equal to $L_0$ (See \cite{randomize}).

From the experimental point of view, the signals provided by the
PhRG and by the running key (${\bf K}_0$) define voltage levels to
be applied by the PCI to the optical modulator OM. Each voltage
$V_k$ generated is associated with a specific basis of the $M$-ry
scheme.
 The pulsed mesoscopic coherent state at the
input (see Fig. \ref{fig1}) can be seen as a linearly polarized
state of light. {\em Orthogonal} polarizations define bits 0 or 1.
The input pulse is modified by the action of the OM into a state
(e.g., elliptically polarized light) ${\bf Y}({\bf R},{\bf K}_0)$
that is sent to B. Without the modulation given by $V_k$ the
output signal would show the sequence ${\bf R}$ of orthogonally
linearly polarized  states (bits 0 and 1) on a single basis.  The
$V_k$ modulation converts these signals to a {\em non-orthogonal}
set of $M$-ry states. A similar line of reasoning applies to phase
modulated signals, where phases 0 and $\pi$ provide the two bits.

{{\bf 4.} \em By knowing the sequence of bits ${\bf K}_0$, Bob
demodulates the received
sequence obtaining ${\bf R}_1$.} \\
At the receiving station, by applying the shared key ${\bf K}_0$
Bob {\em demodulates}
 the changes introduced by A  and
reads  the resulting true random stream ${\bf R}_1$  of
orthogonally polarized light states. A and B now share a fresh
sequence of random bits ${\bf R}_1$.

{{\bf 5.} \em  Bob obtains a fresh random sequence ${\bf R}_2$
from his PhRG and sends it to A, ciphering the sequence in blocks
of size $K_M$. Ciphering bits are taken
from the earlier sequence received ${\bf R}_1$}.\\
Each sequence of bits, of length $K_M$, from ${\bf R}_1$ define
the ciphering basis for $K_M$ fresh bits in ${\bf R}_2$.
 By knowing ${\bf R}_1$, A reads ${\bf R}_2$  with
perfection. The first cycle is complete.

{\bf 6.} {\em A and B continue to exchange random sequences as described in the first cycle}.\\
Subsequent cycles can be performed and in each cycle, blocks of size $K_M$ are ciphered to keep the total length in each
cycle constant and equal to $L_0$. A and B can then share sequences of random bits obtained from the PhRGs. A shared random
sequence can be used to re-start a cycle by A or B whenever an interruption occurs.

{\bf 7.} {\em A and B apply information reconciliation and
 privacy amplification to distill a final sequence of bits}.\\
The process of privacy amplification discards bits in the sequence and, consequently, destroys the short-ranged bit-cipher
correlations due to the block ciphering. As the PhRGs present no bit correlation, the final shared random sequence will
present a similar statistical property.

These steps describe the protocol without discussing security
aspects. However, being a physical protocol, it would be
incomplete without specifying $\langle n \rangle$ and $M$. These
parameters have to be provided, under the  initial conditions
presented, and a quantitative {\em measure} of the security level
associated with them has to be derived. This is the subject of the
following sections.

\section{Bit encoding and the physical protecting mechanism}

The physical protecting mechanism in this case is the same as that
on which the $\alpha \eta$ systems are based. Although it has
already been described, with examples, in Ref. \cite{ykb}, it will
be presented here and discussed to clarify the security provided
by the quantum noise of light to this key distribution system. A
bit-by-bit proof \cite{qcmc02}, based on a Positive Operator
Valued Measured theory (POVM), will follow. The choice of a POVM
demonstration relies on its generality once the wave function or
the  density matrix that represents the physical process is
chosen. The resulting analysis carries the information content in
the density matrix and has broad validity.  This is particularly
useful because the eavesdropper should be allowed to use any
technology or attack (beam-splitter, cloning, homodyne
measurements and so on) and a general protection cannot be based
on particular threat models.

\begin{figure}
\centerline{\scalebox{0.4}{\includegraphics{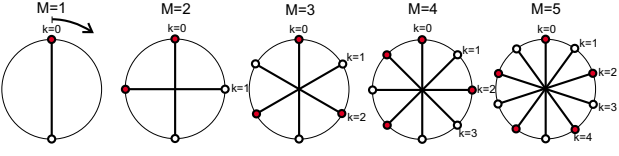}}}
\caption{\label{fig2} Ciphering wheels for phase angles $\phi_k$. Cases $M=1$ to $M=5$ are shown for a given bit.
 Each $k$ value specifies a two-state basis $(0,1)$ where component states are separated by $\Delta \phi=\pi. $ }
\end{figure}

The security analysis to be presented covers both polarization and
phase modulation of optical signals. In the case of free-space
implementation, the coherent states defining each bit are  two
orthogonal modes of polarization. In the phase ciphering, two
modes separated by a phase of $\pi$ could be used. In the {\em
polarization} case the running key specifies a polarization basis
from a set of $M$ uniformly spaced two-mode bases spanning a great
circle on the Poincar\'e sphere. Fig. \ref{fig2} sketches the
$M$-ry ciphering protocol as implemented  in the $\alpha \eta$
systems \cite{ykb,qcmc02} where closest bits are mostly distinct
from each other. In this key distribution  scheme, the same $M$-ry
scheme is utilized.
Each basis represents a polarization state {\em and} its antipodal
state at an angle $\pi$ from it (bits 0 and 1). The mapping of the
stream of bits onto  points of the Poincar\'e sphere is the {\em
key} to be shared by A and B. It points precisely to the basis
being used at each bit emission. Each $k$-basis is defined by the
Poincar\'e angles $\Theta_k$ and $\Phi_k$.
 The number of bases $M$ chosen should be such that the uncertainties caused by the quantum noise of light on
  the polarization angles leads to a large error. This can be understood in a variety of ways; for example,
 by directly writing the manifold of two-state $\{|\Psi(\Theta_k,\Phi_k)\rangle\}$ bases  in Cartesian $(x,y)$
 coordinates fixed at the OM physical axes (chosen at 45$^0$ from the horizontal) gives
\bea
\label{poinc2}
|\Psi(\Theta_k,\Phi_k)\rangle= |\alpha \gamma(\Theta_k,\Phi_k)\rangle_x \otimes |\alpha \delta(\Theta_k,\Phi_k)\rangle_y \:\:,
\eea
where $\alpha$ is the coherent amplitude and $\gamma$ and $\delta$ are the projections on $x$ and $y$.
\bea
\gamma&\!=&\!\!\left[(1-i)e^{i \Phi_k/2}\cos(\Theta_k/2) + (1+i)e^{-i \Phi_k/2}\sin(\Theta_k/2) \right]
\!,\nonumber\\
\delta&\!=&\!\!\left[(1+i)e^{i \Phi_k/2}\cos(\Theta_k/2) + (1-i)e^{-i \Phi_k/2}\sin(\Theta_k/2) \right]\!.
\nonumber
\eea
For example, on a great circle set by $\Theta_k=\Theta_p=\Theta_0$,
the overlap $\langle \Psi(\Theta_k,\Phi_k)|\Psi(\Theta_p,\Phi_p) \rangle$ between states $k$ ($\Phi_k=\frac{ \pi}{M}k$) and
$p$ ($\Phi_p=\frac{ \pi}{M}p$) gives
\bea
\label{overlap}
|\langle \Psi(\Phi_k)|\Psi(\Phi_p)\rangle|^2=e^{-2\langle n \rangle\left[ 1-\cos\left( \frac{\Phi_p-\Phi_k}{2} \right) \right]}.
\eea
This will define the polarization angle {\em uncertainty} produced by the shot noise associated with the coherent states.
For large $\langle n \rangle$ the periodic functions in Eq. (\ref{overlap}) can be expanded around $\Phi_p$,
as $\Phi_k\simeq\Phi_p+\Delta \Phi$,  giving $|\langle \Psi(\Theta_k,\Phi_k)|\Psi(\Theta_p,\Phi_p) \rangle|^2\simeq
\exp\left[-\Delta \Phi^2/(2  \sigma^2)\right]$.
$\sigma^2=1/\langle n \rangle$ is the uncertainty associated with
the Poincar\'e' angle. This uncertainty is directly associated
with light's shot noise and cannot be overcome regardless of one's
precision capabilities.
 Without knowing the precise basis sent (or angle), E cannot obtain the bit that is sent.
Her  measurement of the polarization angle becomes uncertain by
the uncorrelated noise \cite{ykb,qcmc02} in the two axes
 ( $\langle n_1 n_2\rangle=\langle n_1\rangle\langle n_2\rangle$ ).
It will be shown that this noise can be used judiciously to
prevent an eavesdropper from accessing the information while the
legitimate receiver B can control it. This access is given by the
knowledge of the key: the legitimate receiver projects the
received signal completely onto one of the physical axes of the
receiving system (e.g. the PBS in Fig. \ref{fig1}) and this way
the associated noise becomes irrelevant to his binary
determination (See Refs. \cite{ykb,qcmc02} for experimental
results). Receiver B can even support moderate misalignments of
his bases system because whenever most of the light falls into one
of his detectors this would indicate the correct bit. In contrast,
for Eve, apart from the uncertainty caused by the noise, even a
small misalignment will give  her an incorrect  basis.
Furthermore, her measurement system needs high resolution and
precision to obtain reliable data for analysis. The number of
bases $N_\sigma$ within $\sigma$ is $N_\sigma = M \sigma/ \pi=M/(
\pi \sqrt{\langle n \rangle})$. The system should be designed, as
it will be shown, such that $N_\sigma$ covers a reasonable number
of adjacent bases.

{\em Phase} modulation of the signals can be utilized by creating
two pulses delayed by a fixed amount of time and introducing a
phase difference $\phi_b$ between them to represent bits 0 or 1
(e.g., $\phi_b=0$ and $\pi$). An extra phase difference $\phi_k$
is provided by the $K_M$ shared bits. At the receiver, these
pulses can be made to interfere and by subtracting the phase
$\phi_k$, B can recover each random bit sent. Formally, this
 phase encoding could be written starting from a  coherent state
 $|\alpha\>$ that is split into  a
two-mode coherent state $|\Psi_0 \>=|\alpha/\sqrt{2}\>_1 \otimes |\alpha/\sqrt{2}\>_2$. Bit encoding using the two-mode
state, represented by annihilation operators $a_1$ and $a_2$, can be done by \bea \label{Psi_b} |\Psi_b \>=e^{- i J_z
\varphi_b}|\Psi_0 \>=|e^{- i \varphi_b/2}\frac{\alpha}{\sqrt{2}}\>_1 \otimes |e^{ i \varphi_b/2}\frac{\alpha}{\sqrt{2}}\>_2,
\eea where $J_z=\left( a_1^{\dagger} a_1 -a_2^{\dagger} a_2  \right)/2$. This phase modulation can also be interpreted as a
relative one, with the zero reference taken at one of the states
. A crucial ingredient in the security
demonstration is that the modulation operations have to be unitary or energy conserving. In this way, the input energy
associated to each pulse will have to be distributed between the two modes.
 Precise information about the energy content in each mode is not
needed, but one is assured that all energy is being accounted for
in the demonstration. Although losses are unavoidable in real
systems, this condition also reflects the fact that technical
losses are expected to decrease with advances in technology and so
they can be considered asymptotically negligible.  Therefore, for
a modulation system that is not energy conserving in principle,
the following demonstration does not apply.

 In the  phase modulation case, one can associate an index $\nu$, in general,
  to the ciphering angle $\phi_{\nu}$ to represent a possible
 applied modulation. This index $\nu$ could represent a discrete or a
 continuous variable determined by a general distribution. In this
 $M$-ry scheme $\nu=k$.

 A ciphered bit in the two-mode state will be written
\bea
\label{Psi}
&&|\Psi_{b\nu} \>=e^{- i J_z (\varphi_b+\phi_\nu) }|\Psi_0 \>\nonumber \\
&&=|    e^{- i (\varphi_b+\phi_\nu)/2 } \alpha/\sqrt{2}\>_1
\otimes |    e^{  i (\varphi_b+\phi_\nu)/2 }
\alpha/\sqrt{2}\>_2\:\:, \eea where $\varphi_b\:(=0, \:\pi)$
specifies the bit being ``sent'' and $\phi_\nu$ is the ciphering
phase. The overlap of $| \Psi_{b\nu} \>$ and $|\Psi_{b\mu} \>$
leads to an equation similar to (\ref{overlap}).

\section{Eavesdropper's minimum probability of error}

To show that this key distribution scheme is secure {\em two} basic
points have to be demonstrated:\\
1)  For a fresh bit sent,
the minimum probability of error $P_e^E$ that an eavesdropper can
achieve in the bit determination
must be guaranteed to be {\em arbitrarily} close to $1/2$, \\
 2) The use of a given random sequence  $2 \times K_M$, one time as a
 ``message'' and  the second time as a
 cipher for the fresh random sequence, still allows one to set
 $P_e^E \rightarrow 1/2$.

As a starting point for the {\em first} part of the demonstration,
the density matrix $\rho$ for all
possible two-mode states resulting from ciphering a bit $b$ is written as
\bea
\rho_{b}=\frac{1}{L}  \int_0^{L}  P_{\phi_{\nu}}
|\Psi_{b\nu} \> \<  \Psi_{b\nu} | d\nu\:\:,
\eea
where $L$ is the space spanned by $\nu$ and $P_{\phi_{\nu}}$ describes
a general  phase distribution.
The optimal POVM for discriminating between $\rho_0$ and $\rho_1$
(or $\Delta\rho=\rho_1-\rho_0$),
in the polarization case, was first applied in Ref~\cite{qcmc02}.

Calling $\Pi_1$ and $\Pi_0$ ($\Pi_1+\Pi_0=$I) the projectors over
eigenstates with the positive and negative eigenvalues of $\Delta
\rho$, the probability of error $P_e^E$ is \bea \label{P_e}
P_e^E={\rm Tr}[p_1 \Pi_0\rho_1+p_0 \Pi_1\rho_0] \:\:, \eea where
$p_1$ and $p_0$ are {\em a-priori} probabilities to find a state
in $\rho_1$ or $\rho_0$, respectively. $P_e^E$ defines the minimum
probability of error that is caused by a wrong choice of bases by
Eve when she tries to determine a bit sent. Of course,  error
levels higher than the one given by Eq. (\ref{P_e}) can be found
but the interest here is to find Eve's lower bound of error in a
bit-by-bit determination.

$P(\phi_{\nu=k})$  randomly establishes the index $k$ associated
with discrete phase values $\phi_{k}$ in the ciphering wheel shown
in Fig \ref{fig2} where adjacent bits to a given $k$ are mostly
distinct bits from the $k^{\mbox{th}}$ bit. For this
implementation the location of the two-state bases are given by
\begin{eqnarray}
\label{angle}
\phi_{k}=
 \pi   \left[\frac{k}{M} +
\frac{1-(-1)^k}{2}  \right]\:\:\:,
\:\:\:k=0,1, ...,M-1\:\:\:.
\end{eqnarray}
For equal a-priori probabilities $p_1=p_0=1/2$,  Eq. (\ref{P_e})
reduces to \bea \label{perr} P_e^E=\frac{1}{2}{\rm
Tr}[\Pi_0\rho_1+\Pi_1\rho_0]=\frac{1}{2}\left(1-{\rm Tr}
[\Pi_1\Delta\rho]\right)\nonumber\\=\frac{1}{2}(1-2 \sum_j
\lambda_j) \:\:, \eea where $\lambda_j$ are the positive
eigenvalues to be obtained from
\bea \label{delta_rho} \Delta
\rho=\!\!\frac{1}{M}\sum_{\nu=k=0}^{M-1}
  e^{-i J_z \phi_\nu}
            \left(  |\Psi_1\> \< \Psi_1| - |\Psi_0\> \< \Psi_0|\right) e^{i J_z
            \phi_\nu}.
\eea Eq. (\ref{delta_rho})  can be expanded as \bea \Delta
\rho=\sum_{q=-\infty}^\infty \sum_{q'=-\infty}^\infty \Delta
\rho_{q,q'} | \Phi_{q}\>\>  \<\< \Phi_{q'}|\:\:, \eea where \bea
\label{deltarho} \Delta \rho_{q,q'}=-2 i e^{-|\alpha|^2}
\sqrt{I_{2|q|}\left( |\alpha|^2 \right) I_{2|q'|}\left( |\alpha|^2
\right)}
\times\nonumber\\
 \sin\left[(q'-q)\pi/2  \right]
 e^{i(q'-q)\pi/2}
\frac{1}{M}\sum_{k=0}^{M-1} e^{i \phi_k (q'-q)} \:\:, \eea where
$I_{j}$ is a Modified Bessel function, \bea \:\:\:|
\Phi_{q}\>\>\!=\!\frac{1}{\sqrt{I_{2|q|}(|\alpha|^2)}}
\!\!\sum_{J=|q|}^{\infty} \! \frac{   \left(
\alpha/\sqrt{2}\right)^{2 J}   }{\sqrt{ \left(J-q\right)!
\left(J+q\right)! }}
| J,q\>\> \\
 \mbox{and}\:\:\:| J,q\>\>=|J-q\> \otimes |J+q\>\:\:.
\eea
 From the positive eigenvalues of Eq. (\ref{deltarho}),  the minimum probability of error,
 Eq. (\ref{perr}), can be calculated.

 Eq. (\ref{delta_rho}) can be expanded in several ways,  the
 adopted
 expansion uses the angular momentum basis $| J,q\>\>=|J-q\> \otimes
 |J+q\>$, that is a natural basis to deal with angular rotations.

\begin{figure}
\centerline{\scalebox{0.45}{\includegraphics{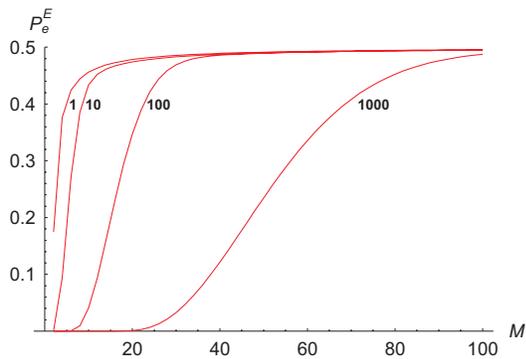}}}
\caption{\label{fig3} $P_e^E$ as a function of $M$ for
$|\alpha|^2\equiv \< n \>=1,\: 10,\:100$, and $1000$. }
\end{figure}

Assuming that $k$ values have uniform probability of occurrence one can show that the number of
a-priori probabilities for the number of occurrence of even-$k$ or odd-$k$ lines, given $M$, is
\begin{eqnarray}
p_{even-k}(M)=\frac{1-(-1)^M+2 M}{4 M}\:\:\:,\nonumber\\
p_{odd-k}(M)=\frac{-1+(-1)^M+2 M}{4 M}\:\:.
\end{eqnarray}
For simplicity and without loss of generality, let us adopt bases even in $M$, where $p_{even-k}(M)=
p_{odd-k}(M)=1/2=p_1=p_2$, to show numerical examples.
Figure~\ref{fig3} shows the minimum probability of error as a function of the number of ciphering levels $M$.
$P_e^E$ goes very fast to the asymptotic pure-guessing limit of $1/2$ as $M$ increases.
It is then shown that
the minimum probability of error $P_e^E\rightarrow 1/2$, at a fixed average number of
photons $|\alpha|^2$, can be achieved by increasing the number of bases $M$ adequately.
 This demonstrates that in this scheme an eavesdropper {\em cannot}
 obtain the individual bits sent, regardless the precision of her devices.
 The physical origin causing this impossibility for Eve to obtain an angle $\phi_k$ and therefore the associated bit,
 rests in the source emission itself. Although a deterministic angle can be applied to the modulator the resulting light output
 is probabilistic in nature and in single shot measurements with mesoscopic states the resulting field does {\em not} carry
 this applied angle $\phi_k$ but presents a distinct $\phi_k^{\prime}$. This is a Nature's fact that cannot be changed regardless
 the measurement applied (homodyne, heterodyne, and so on).
This completes the first part of the demonstration.

For the {\em second} part of the demonstration, one has to show how repetitions of the cipher to encode the distinct bits
generated in the random process  increase the resolution achievable by the eavesdropper over the signal  sent. One should
recall that cipher repetition are used in the block ciphering described in step 3 of the protocol. As discussed in
\cite{randomize}, these repetitions are {\em not} necessary because the ciphering procedure can be randomized bit-by-bit
through use of a stream cipher using the $K_M$ sequences as seed keys.

One could ignore this perfectly possible randomization an calculate a much more drastic case, one where both cipher and bit
were repeated $r$-times. This can be seen as an overestimated upper bound for the actual situation because each random bit
sent is a fresh bit in the sequence being sent. As photon numbers in distinct coherent pulses of same amplitude fluctuates
in an uncorrelated way, in $r$-repetitions of a signal the resolution achieved for extraction of this signal increases with
the number of repetitions. This is quantitatively obtained from the $r$-product of Eq. (\ref{overlap}): $P(r;k|p)\equiv
P(k|p)^{\otimes r}\simeq
\exp\left[-r \Delta \Phi^2/(2  \sigma^2)\right]$. This Gaussian
process gives the standard deviation $\sigma'=\sigma/\sqrt{r}$
associated with the angle uncertainty in a measurement process.
This uncertainty $\sigma'$ is equivalent to the one obtained from
a {\em single} shot measurement with the photon number $r \langle
n \rangle$.
 In other words, a single shot using $r$-times the laser power
 will give the same signal resolution for a bit reading as the
$r$-repeated sequence with $\langle n \rangle$.
 Consequently, for a fixed $M$, the $r-$repetition of the
 random sequence then reduces $P_e^E$ from
 $P_e^E(\langle n \rangle)$ to $P_e^E(r \langle n \rangle)$.
The dependence of $P_e^E$ can be calculated as a function of $\langle n \rangle$ and $M$ for arbitrary numbers. Therefore,
the system can be designed to a desired security level $P_e^E$, through the correct choice of $\langle n \rangle$ and $M$.
As a numerical example consider, say,  $M=32$ (or $K_M=5$ bits) with $\langle n \rangle=100$ to achieve $P_e^E=0.476$ in a
single shot (see Fig. \ref{fig3}). To guarantee the same security  level ($P_e^E=0.476$), due to the $r=2\times K_M=10$
repetitions, one should use $M=90$ ($K_M\sim 7$) corresponding to  $\langle n \rangle=10\times 100=1000$.   The conclusion
is  general regardless of the specific numerical example. Proper scaling can be done for other intensity levels adequate for
the sensitivity of the detection system. Although this is an overestimated calculation it is adequate for our purposes to
show that the protection level can be increased according it is needed.  The alternate encryption described in
\cite{randomize} reduces this overhead substantially because each level used to cipher each bit is close to have occurred in
a truly random.

It has been shown that the transmission stages A$\rightarrow$B and B$\rightarrow$A can be made secure under individual bit
attacks. The quantity $P_e^E$ can be connected to the bit-error-rate probability and
 entropy measurements such as mutual information or relative entropy,
 can be directly derived from it.

The following section discusses some aspects of attacks on this scheme. As there are no ciphertext or known-plaintext
involved in the transmission of random bits, attacks on the transmitted random sequences to obtain the key $K_0$ have to
start considering a guessing probability of $1/2$ for each bit or $1/2^{L_0}$ for any complete sequence sent. Considering
that for each new bit sent the random noise may produces $\sim N_{\sigma}$ possible outcomes in a measurement process, the
number of possibilities to be considered grows {\em exponentially} depending on $L_0$ and $N_{\sigma}!$. This indicates that
an unsurmountable computational problem would occur for a large number of bits sent.

\section{Eve's record,
correlations and drawbacks}

Next, one has to show that the security level calculated also
holds under the basic conditions already defined: I) The
eavesdropper is allowed to have full access to the random signal
sequence being generated. II) She would work near the source to
avoid signal losses.

\subsection{Bit-cipher block correlations}

 In general, any degree of correlation in a bit sequence can be explored by Eve to decrease her degree of
 uncertainty. While use of true random numbers eliminate intrinsic
 correlations in the generation process the $K_M$ block
 cipher utilized introduces a short range bit-cipher correlation. The increased number
 of levels $M$ utilized was aimed to increase Eve's bit-by-bit probability of
 error to the guessing levels even under the block ciphering used. Although Eve cannot obtain the
 random bits sent from the measured signals one may argue that
 some correlation may be detected due to the bit-cipher
 correlation.
 As one can see from the given numerical examples,
 the condition $N_\sigma > K_M$ can be easily applied to the communication scheme with mesoscopic states.
 This condition assures that the uncertainty produced by the noise overcomes the
 amount of knowledge associated to these correlations. In a linear algebraic system it
 corresponds to a number of unknowns larger than the number of
 available equations. Nevertheless,
 {\em information reconciliation} and {\em privacy
amplification} \cite{cachin} distill a random sequence about which Eve has a negligible amount of information and also has
the additional effect of destroying the short bit-cipher correlations.
 Basically, the fact
that the information known by the legitimate parties {\em differs} from that obtained by the eavesdropper is what allows A
and B to achieve the secrecy goal.


\subsection{Homodyne and delay line}

Homodyne and heterodyne techniques can be also utilized by Eve  to
obtain the signals with better precision than a direct detection
measurement. With knowledge of the key sequence Bob and Alice
always utilize the proper quantum measurement basis for their
optimal {\em binary} detection, assuring a resolution superior to
the one obtained by Eve. In principle, if the seed key is
available to Eve, after her records were created, she could apply
rotations over the recorded signals to obtain the correct
sequence. This has to be performed in a sequential operation  due
to the lack of structure in the PhRGs generation of random
numbers. This differs from performing a mathematical operation
with a deterministic function to obtain a result $x_{n-1}$ given
$x_{n}$, up to $x_0$. Nevertheless, assuming that this inversion
is possible in principle, Eve could obtain the random sequence
once the starting key is available \cite{eric}. This will be
equivalent to a futuristic optical delay line that could be tapped
on demand and Eve could wait as long as necessary until the shared
starting key is made available to her and only then perform her
measurements perfectly mimicking Bob's measurement over her copy
of the signals. With the key, E does not need the same resolution
as before and the applied rotations would lead her close enough to
the correct axis orientation and to bit identification.
 Therefore, {\em the shared starting key has to be protected at all times}.
 This is the fundamental drawback of this system \cite{key disclosure},
 although it is not important in the current state of technology .
 In essence, assuming a protected starting key,
 this method can be described as a one-time-pad extender.

 The futuristic delay line discussed led to the protection of the
 starting key at all-times because there are no in-principle impediments for
 creation of such a device. One could also invoke a perfect photon
 number cloning so that Eve could obtain multiple perfect copies
 of each signal sent. Perfect copies would allow Eve to obtain the correct bit
 sequence but perfect cloning violates the no-cloning theorem.
  Realistic amplification schemes with linear optical
 amplifiers introduce spontaneous noise that decreases Eve's
 signal resolution. Imperfect cloning \cite{lutkenhaus} using realistic number
 distribution consisting of mixtures of stimulated and spontaneous thermal
 processes are akin to linear optical amplification and
 degrades signal resolution.


\subsection{Phase measurement bounds}

Even without reference to a particular setup, it is worth
comparing the obtained numerical results with the bounds on phase
measurements imposed by the smallest detectable phase shifts
usually considered (See \cite{barnett-fabre-maitre} and references
therein). They are: 1) the standard quantum limit ($SQL$), given
by $\Delta \phi_{SQL}\sim1/\sqrt{N}$, 2) Squeezed light limit
($Sqz$), given by $\Delta \phi_{Sqz}\sim1/N^{3/4}$, and 3)
``Heisenberg'' limit $\Delta \phi_{H}\sim1/N$. A homodyne setup to
identify a given phase that uses, say, a squeezed field has to
scan the phase interval of interest to probe for the best
resolution region where the phase measurements lapse below shot
noise condition. In the case of interest in this paper, each
signal pulse will be set randomly in one basis chosen among $M$
bases. The minimum interval between bases is $\Delta
\phi_{min}=\pi/M$.  A uniform scan on the semicircle ($\pi$) with
maximum efficiency will find the particular basis sent within a
fraction $\Delta \phi_{min}/\pi=1/M$ of the pulse containing $N$
photons. One could therefore impose the condition for
indistinguishability of the sent phase using the lowest limit
given by $\Delta \phi_{H}$: \bea \Delta \phi_{H} > \Delta
\phi_{min} \:\:. \eea Writing $\Delta \phi_{H}$ to take into
account the optimized fraction of the pulse, one has \bea \Delta
\phi_{H}=\frac{1}{(1/M)N}\:\:, \eea and therefore \bea M >
\sqrt{\pi N}\:\:. \eea This condition is satisfied under the
conditions exemplified in Fig. \ref{fig3}, in agreement with
obtained results.

\subsection{Speed costs and other aspects}

The main cost for the  security obtained in this  scheme is the $M$-ry number of levels needed. An increased number of
levels demands an increased number of bits to provide the necessary resolution and a wider dynamic range for the waveform
generators to provide such modulations at high speeds. For example, in case one assumes the overestimated $2 K_M$ repetition
this will decrease the bit output rate from, say, 10GHz, to 10GHz/$2 K_M$. For a nonrepeated cipher, if one uses $M=1000$
($K_M \sim10$), and $\langle n \rangle=10^4$ the number of levels covered $N_\sigma=M/(\pi \sqrt{\langle n \rangle})\sim
3.18$. To keep the same  number of levels $N_\sigma$ covered under the $2 \times K_M$ repetition, equivalent to a single
shot  with the intensity $2 \times K_M$ higher, the number of levels necessary is $M_{new}\sim 4472$. This reduces the speed
to 10GHz/$2 K_{M_{new}}=0.4$GHz from 10GHz. Although this large increase in the number of levels is not necessary
\cite{randomize} it emphasizes the overhead involved when one increases the number of levels $M$.

Exploratory physical attacks by the eavesdropper, such as injection of a strong signal to detect the weak reflections
from the surface of the OM modulator, and from these to obtain the modulation applied, can be easily detected by signal splitting.
 Specific analysis for the physical attacks could be applied  on a case-by-case basis.
The robustness of the signals under signal jamming by an enemy,
for example,  may be of interest for some applications. In this
case,  one  could superpose on the ciphering levels phase and
amplitude modulations and even utilize emission at distinct
wavelengths to provide a set of conditions that the legitimate
parties could use to extract the signals. Again, specific issues
require case-by-case responses. These questions are not related to
the security aspects of interest here. In the same way, computer
attacks \cite{schneier} and general attacks outside the optical
channel are not the focus this work.

\subsection{Signal amplification}

The presented key distribution scheme aims to defeat Eve's actions
at the source, where no losses occurred. Amplification processes
always degrade signal resolution for Eve or Bob.  As Eve is
already defeated at the source, she cannot obtain any improvement
through amplification. On the contrary, the knowledge of the key
allows Bob to distinguish signals as long as he has a good
signal-to-noise ratio. Therefore, amplification is possible for
the legitimate receivers
 because they need a smaller resolution degree than Eve.
 It works as long as A and B can identify signals in orthogonal bases.
 A and B apply simple binary decisions to distinguish between orthogonal bases while Eve, on
 the other hand, needs high resolution to distinguish between
 adjacent levels of the $M$-ry scheme. This is why an increased
 noise for A and B due to the amplification process does not have the same effect
 on Eve's measurement. With the signal protected at the source, noise created by
 amplifiers can be acceptable for A and B until their error bit rate exceeds
 a toleration level.

 Numerical simulations indicate that A and B can utilize
 amplifiers up to distances of $\sim 500$km before signal
 regeneration becomes necessary. These simulations consider the
 spontaneous decay in amplifiers onto the mode to be amplified as
 well as onto the mode orthogonal to the carrier. These added
 noise sources degrade signal resolution for the users and
 increase the bit error rate. Other error sources exist such as
 acoustic and thermal fluctuations but they occur in a much
 slower time scale.

 It is important to observe that any advantage obtained by A and B over E, at the source,
 leads to an increased communication distance for A and B. Although B can support losses as long as this system presents a
 low bit error rate, his superiority over Eve decreases and ceases at the moment where Eve can have an equal or larger
 amount of information than Bob \cite{horace}. These differences can be estimated by calculating
 $P_e^E (0)/P_e^B$, where $P_e^E (0)$ is Eve's error probability at the source and $P_e^B$ is calculated after the amplification stages.
 Ideally, $P_e^E (0)/P_e^B \gg 1$.
 Therefore, any randomization that can be further introduced by Alice at the source
 amplifying Eve's uncertainty will reflect as an increased range for secure communication \cite{horace}. Randomization can
 be increased by several means;  however, increasing the randomization level usually has a cost associated to
 the process that has to be weighted with respect to the gain to be achieved.

Summarizing,  signal amplification is essential for Internet and this key distribution scheme can be tailored for some of
these applications.

 \section{Mutual Information}
\begin{figure}
\centerline{\scalebox{0.45}{\includegraphics{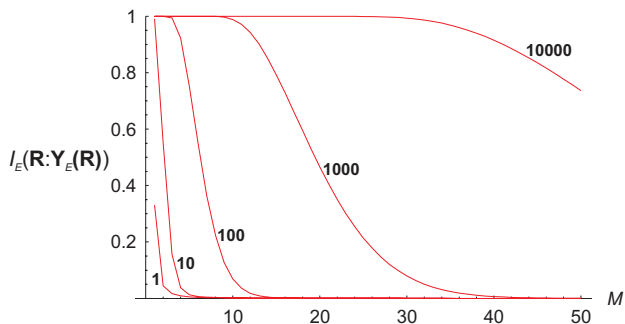}}} \caption{\label{mutual_I} $I_E(R:Y_E(R)$ as a function of $M$
for $|\alpha|^2\equiv \< n \>=1,\: 10,\:100$, and $1000$. }
\end{figure}

The concept of mutual information $I(X:Y(X))$, concerning the amount of information on $X$ giving the observable Y, allows
one  to extract basic information on this key distribution system. The minimum probability of error for Eve, bit-by-bit,
derived in Eq. (\ref{P_e}), $P_e^E$, gives the bit-error-rate for Eve in the binary entropy $H(R|Y_e(R))$. The mutual
information $I_E(R:Y_e(R))=H(R)-H(R|Y_e(R))=1+P_e^E \log_2P_e^E+(1-P_e^E)\log_2(1-P_e^E)$ describes the amount of
information Eve could obtain in a bit-by-bit attack on $R$. Fig. \ref{mutual_I} shows this dependence as a function of $M$
and some values of $\langle n \rangle$. $H(R)=1/$per bit (or $L_0$ for a perfect random stream of length $L_0$).
Although an individual bit-by-bit attack is only a limited attack from Eve it shows the increasingly difficulty the attacker
finds as a function of the variables used.

The presented key distribution process intends to perform a distribution between two users of a fresh sequence of random
numbers generated by one or more PhRG's. While the final shared sequence is truly random, as it was thoroughly discussed,
each sequence was obtained using the former sequence to cipher it. The whole process was then dependent on the secrecy of
the fist key sequence $K_0$ (of length $L_0$).  As long as one chain in the sequence is not compromised, the whole sequence
results secure. In other words, although A and B may share an arbitrarily long sequence of random numbers, the security of
the process relies in a single element of the chain. This is distinct from processes described as {\em fresh key generation}
in the conventional nomenclature where one aims to create keys that are unconditionally secure even after the starting key
is compromised. This goal has not yet been achieved outside from single photon protocols and it is a constant matter of
study \cite{horace}.

This intuitive dependence of the security of the chained sequence of keys on a single element of the chain can be formalized
as follows. Using the definition of mutual information utilized above, \bea I(R:Y(R))=H(R)-H(R|Y(R))\:,\eea and writing the
difference between the mutual informations between B and E at each cycle of length $L_0$ : \bea \Delta
I&=&I_B-I_E\nonumber\\&=&[H(R)-H(R|Y_B(R))]-[H(R)-H(R|Y_E(R))]\nonumber\\&=&H(R|Y_E(R))-H(R|Y_B(R))\:.\eea From the fact
that, at best, $H(R|Y_B(R))=0$ and $H(R|Y_E(R))=L_0$ in the first cycle, one  sees that $\Delta I \leq L_0$, what shows that
the uncertainty may be kept at the first sequence $L_0$ with a level given by $L_0$ itself. The conditions that Bob reads
without errors the random numbers sent ($H(R|Y_B(R))=0$) in a sequence of length $L_0$ and that the secrecy of the shared
key for Eve is perfect ($H(R|Y_E(R))=L_0$) are basic and reasonable assumptions.  To complete the derivation, one may write
the chaining rule \cite{nielsen_chuang} $H(L_1,L_2,\cdots,L_n|Y_E)=\sum_{i=1}^nH(L_i|Y,L_1,\cdots,L_{i-1})$ to see
that \bea &&H(L_1,L_2,\cdots,L_n,\cdots|Y_E)=\nonumber \hspace{4cm}\\
&&H(L_1|Y_E)+H(L_2|Y_E,L_1)+H(L_3|Y_E,L_1,L_2)+\cdots\eea One can see from this rule that in the first round where $L_1$ (or
$\{{\bf R}_1,\cdots {\bf R}_{K_M}\}$) is ciphered with $K_0$, unknown to Eve, Eve's entropy knowing $Y_E$ is
$H(L_1|Y_E)=K_0$, as long as no information on $K_0$ is leaked to her. However, in case Eve knows $L_1$ and $Y_E$ she could
know $L_2$, that is to say $H(L_2|Y_E,L_1)=0$ because this provides her with the same information that Bob has, and so on
with the subsequent terms. Therefore $\Delta I=K_0$ or, in other words,  Eve's uncertainty is $K_0$ (length $L_0$). In case
Eve is unable to obtain $L_1$, $L_2 \cdots$ this result does not apply, of course, and Eve's knowledge on the whole sequence
of transmitted random numbers is null.  This analysis stands for any length $L_0$, that has to be tailored according one
needs. For $L_0=10^9$, as an example, implies that to break the sequence Eve has to order correctly $10^9$ bits, what gives
her a probability for success of $1/(2^{10^9})$. Privacy amplification steps further decreases $P_e^E$ as well as destroys
short range correlations in original random streams. Even periodic replacements of the starting keys can be introduced and
the whole process can be tailored to any advances in computational capabilities.

\subsection{Quantum noise}

One should understand that the inherent quantum noise in the system sometimes invalidates conventional techniques one could
use in a noiseless system. For example, in the case of a classical $M$-ry ciphering, or ciphering with intense coherent
signals, Alice could send repeatedly a given bit $r$ to Bob fixed in any basis of the $M$-ry detection system. Bob measures
successively $r_1=r, \:r_2=r, \:r_3=r, \cdots$. The following properties then holds: 1) Any pair of $r$ obeys $r \oplus
r=0$. Therefore, 2) If a message bit $x$ is ciphered as $y=x \oplus r$ by Alice, Bob is able to recover $x$ using any
obtained $r$ by performing $y \oplus r=x$. This holds  because $r$ is obtained noiseless and, therefore, $r \oplus r=0$ .
Classical one-time pad applications are ``noiseless''. Quite distinctly, when using a mesoscopic coherent state, a
repeatedly sent bit $r$ will be read by a receiver that does not know the basis to be used, as $r_1=r+\Delta_1$, $r_2=r
+\Delta_2$, ..., where $\Delta_j$ express the effect of the noise in the channel. In this case,  1) Any pair of $r$ obeys
$r_i\oplus r_j= \Delta_1+ \Delta_2$ instead of $r \oplus r=0$ ; 2) If a message bit $x$ is ciphered as $y=(x \oplus r)_i$ by
Alice and Bob has $r_j$, he  obtains $y \oplus r_j=x+ \Delta_i+\Delta_j$ instead of $x$. This only stresses that the
knowledge of the bases used to transmit the random sequences or keys is a crucial step for this key distribution system. One
should realize that the channel noise, inherent to the light field, has its source at the emitter itself, ignoring all
technical noises eventually present.


\section{Conclusions}
It has been demonstrated that a fast
 key distribution scheme between two stations can be implemented where
the practical physical limitations are set by the speed of the electro-optic modulators and acquisition electronics
available through current technology.  A secure random sequence obtained can be utilized in ``Vernam's one-time-pad'' sense,
for applications that demand unconditional security. Fundamentally,
 the system allows for signal {\em amplification}, as in the $\alpha \eta$ systems. The system works as long as the receiver
 has a good signal-to-noise ratio after the last amplification
 stage and the occurred losses do not give him an information level worst than Eve's.
 The possibility of amplification paves the way for long
 distance key-distribution protocols protected by the quantum noise of light,
 offering a practical advantage over
 single-photon protocols.

{\bf Acknowledgements --} The author has been supported under grants DARPA/AFRL-F30602-01-2-0528 and NSF-PHY-0219382. The
views and conclusions contained herein should not be interpreted as representing  any endorsement from the cited agencies
and just represents a personal view on this problem.  Thanks are due to Horace P. Yuen for enlightening discussions on
cryptography and to Prem Kumar, sharing his expertise and experimental knowledge in the development of the $\alpha \eta$
systems. G. Mauro D'Ariano and M. G. A. Paris are acknowledged for the joint work on the POVM demonstration for the
polarization case \cite{ykb}. Eric Corndorf is acknowledged for useful discussions.


\begin{thebibliography}{50}
\bibitem{bennett-brassard}
C. H. Bennett and G. Brassard, ``Quantum cryptography: public-key
distribution and coin tossing'',
 Proc. IEEE International Conference on Computers, Systems and Signal Processing, Bangalore, India, pp. 175-179, 1984.

\bibitem{alamos}
N. Gisin, G. Ribordy, W. Tittel, and H. Zbinden, Rev. Mod. Phys.
{\bf 74}, pp. 145-195 (2002). R. J. Hughes, J. E. Nordholt, G. L.
Morgan and C. G. Peterson, QELS Conference, OSA Technical Digest,
Vol. 74, p. 266 (2002).

\bibitem{qrepeater}
H.-J. Briegel, W. D\"ur, J. I. Cirac, and P. Zoller, Phys. Rev.
Lett. {\bf 26}, 5932 (1998). A. Mair, J. Hager, R. L. Walsworth,
and M. D. Lukin, Phys. Rev. A {\bf 65}, 031802R (2002).


\bibitem{private}
H. P. Yuen, in  ``Ultra-secure and Ultra-efficient Quantum Cryptographic Schemes for Optical System, Networks, and the Internet'',
 unpublished, Northwestern University (2000).

\bibitem{ykb}
G. A. Barbosa, E. Corndorf, and P. Kumar, Quantum Electronics and Laser Science Conference, OSA Technical Digest {\bf74},
pp. 189-190 (2002).

\bibitem{qcmc02}
G. A. Barbosa, E. Corndorf, P. Kumar, and H. P. Yuen, Phys. Rev.
Lett. {\bf 90}, 227901 (2003); and also in quant-ph/0212018 v2 21
Apr 2003.
 G. A. Barbosa, E. Corndorf, P. Kumar, H. P. Yuen, G. M. D'Ariano, M. G. A. Paris, and P. Perinotti,
"Secure Communication using Coherent States",
 in The Sixth Int. Conference on Quantum Communication, Measurement and Computing, July 2002 (Rinton Press, Princeton, April 2003), pp. 357-360.

\bibitem{bases}
Although the notation ``$M$-ry'' has been used in the literature to designate either the number of bases used or the number
of states ($=2\times$ number of bases), this paper adopts $M$ to designate the number of bases. A reason for this is that
the number of states attached to each basis can, in fact, be modified at will and may involve, in general, even more than
the two states presented in this paper.

\bibitem{one way}
The protocol for the key distribution scheme can be similarly
developed for a one-way or a two-way channel.

\bibitem{quantph03051}
M. Aspelmeyer, T. Jennewein, and A. Zeilinger, M. Pfennigbauer and
W. Leeb, quant-ph/0305105 v1, 19 May 2003.

\bibitem{zbinden}
There is no physical restriction, of principle, to limit the speed
of an optical randomizer.
 Different principles can be utilized to generate random numbers,
e.g. ``A Stefanov, N Gisin, O Guinnard, L Guinnard, H Zbinden,
quant-ph/9907006-2Jul1999'' show an optical randomizer working at
100kHz utilizing a beam splitter.
Although true random number generators are not easy to be created and their randomness have to rely, practically,  on a
finite sequence of tests, no basic reason forbids their existence.


\bibitem{randomize}
Any amongst $M=2^{K_M}$ two-state bases can be generated with $K_M$ bits. Therefore, in an alternate form,  the first
sequence of length $K_M$ can be feed to a linear feedback shift register (LFSR) as a seed key. The output of the LFSR can
generate $2^{K_M}-1$ keys from this input or $2^{K_M}$ including the seed key. This will create $M$ levels randomly chosen
and each of these levels can be used to cipher the random bits from the PhRG to be sent to B. This process further buries
into the noise the information contained in a given sequence $K_M$. This is a possible process instead of a simple block
ciphering using a single basis given by $K_M$.



\bibitem{cachin}
C. Cachin, U. M. Maurer, Journal of Cryptology {\bf 10}, 97-110
SPR (1997).





\bibitem{eric}
Eric Corndorf, private communication.

\bibitem{key disclosure}
In quantum protocols, e. g. BB84, the knowledge of the bases used
after the data is recorded becomes a useless information. However,
the parties must also share a starting secret sequence for
authentication.


\bibitem{lutkenhaus}
D. Bru\ss, J. Calsamiglia, and N. L\"utkenhaus, Phys. Rev. A {\bf 63}, and N. L\"utkenhaus, Phys. Rev. A, 042308 (2001).

\bibitem{barnett-fabre-maitre}
S. M. Barnett, C. Fabre, and A. Ma\^itre, Eur. Phys. J. D
DOI:10.1140/epjd/e2003-00003-3 (2003).

\bibitem{horace}
H. P. Yuen, arXiv:quant-ph/0311061 v5 8 Mar 2004.

\bibitem{nielsen_chuang}
Michael A. Nielsen and Isaac Chuang, {\em Quantum Computation and Quantum Information}, (Cambridge, 2000).


\bibitem{schneier}
B. Schneier, {\em Secrets \& Lies - Digital Security in a
Networked World}, (Wiley Computer Publishing, NY 2000).

\end{thebibliography}
\end{document}